\documentclass[rnote]{aa} % for the research notes
\usepackage{graphicx}
\usepackage{txfonts}
\usepackage{natbib}
\bibpunct{(}{)}{;}{a}{}{,}
\begin{document}
   \title{2MASS J03105986\,+1648155\,AB --- A new binary at the L/T transition
   \thanks{Based on observations collected at the European Southern Observatory, Paranal, Chile, under program 080.C-0786\,A. This work is partly based on observations made with the NASA/ESA Hubble Space Telescope, obtained at the Space Telescope Science Institute (STScI) and is associated with program GO-10208. STScI is operated by the Association of Universities for Research in Astronomy, Inc., under NASA contract NAS 5-26555.}
   }

   %\subtitle{with possible flux reversal}

   \author{M.B. Stumpf
          \inst{1},          
          W. Brandner\inst{1},  H. Bouy\inst{2}, Th. Henning\inst{1}, S.Hippler\inst{1}
          }

  \offprints{M.B. Stumpf}

   \institute{Max-Planck-Institut f\"ur Astronomie, K\"onigstuhl 17,
              D-69117 Heidelberg, Germany\\
              \email{stumpf@mpia.de}
              \and
              Herschel Science Centre, European Space Astronomy Centre (ESA), P.\,O.\,Box 78, E-28691 Vilanueva de la C\~anada (Madrid), Spain
          }

   \date{Received 23 November 2009 ; accepted 17 March 2010  }

% \abstract{}{}{}{}{} 
% 5 {} token are mandatory
 
  \abstract
  {The transition from the L to the T spectral type of brown dwarfs is marked by a very rapid transition phase, remarkable brightening in the \emph{J}-band and a higher binary frequency. Despite being an active area of inquiry, this transition regime still remains one of the most poorly understood phases of brown dwarf evolution.}
  {
  We resolved the L dwarf 2MASS\,J03105986+1648155 for the first time into two almost equally bright components straddling the L/T transition. Since such a co-eval system with common age and composition provides crucial information of this special transition phase, we monitored the system over $\sim$\,3 years to derive first orbital parameters and dynamical mass estimates, as well as a spectral type determination.}
  {We obtained resolved high angular resolution, near-IR images with HST and the adaptive optics instrument NACO at the VLT including the laser guide star system PARSEC.}
  {Based on two epochs of astrometric data we derive a minimum semi-major axis of 5.2\,$\pm$\,0.8 AU. The assumption of a face-on circular orbit yields an orbital period of 72\,$\pm$\,4 years and a total system mass of $\sim$ 30\,-\,60\,$M_\mathrm{Jup}$. This places the masses of the individual components of the system at the lower end of the mass regime of brown dwarfs. The achieved photometry allowed a first spectral type determination of L9\,$\pm$\,1 for each component. In addition, this seems to be only the fifth resolved L/T transition binary with a flux reversal.  }
   {While ultimate explanations for this effect are still owing, the 2MASS\,J03105986+1648155 system adds an important benchmark object for improving our understanding of this remarkable evolutionary phase of brown dwarfs.
Additionally, the observational results of 2MASS\,J03105986+1648155\,AB derived with the new PARSEC AO system at the VLT show the importance of this technical capability. The updated AO system allows us to significantly extend the sample of brown dwarfs observable with high-resolution from the ground and hence to reveal more of their physical properties.}

               
\titlerunning{2MASS\,0310\,+1648\,AB -- A new binary at the L/T transition}
\authorrunning{M.B. Stumpf et al.}
   \maketitle
%
%________________________________________________________________

\section{Introduction}
\label{LT_transition_intro}
The transition from the L to the T spectral types of brown dwarfs is marked by a dramatic change in their near-IR spectral energy distribution (SED) and atmospheric properties. While this has already been an active area of inquiry, it still remains one of the most poorly understood phases of brown dwarf evolution.  As discussed by e.\,g. \citet{Geballe02}, the late-type L dwarfs are characterized by very red near-IR colors, caused by condensate dust in their photospheres and metal hydrides, as well as CO absorption bands. In contrast, the T dwarfs are characterized by again bluer near-IR colors, due to the appearance of CH$_4$ absorption at 1.65 and 2.2 $\mu$m, stronger H$_2$O absorption and the increasing importance of collision-induced H$_2$ absorption (CIA), as well as relatively dust-free photospheres \citep{Geballe02}.  
This change occurs over a comparatively narrow effective temperature range ($\Delta$\,T$_{\mathrm{eff}} \approx$ 200\,K) around 1500\,-\,1300\,K for near-IR L7\,-\,T3 dwarfs \citep{Goli04_2}, implying a very rapid transition phase. Taking into account the interaction between temperature, gravity, metallicity and the physics of atmospheric dust clouds, this area remains a challenge to theoretical models (for different possible explanations see e.\,g. \citealp{Knapp, Tsuji05, Tsuji99, Marley02, Burrows06, Ackerman01, Burgasser02, Folkes07}). 
%_____________________________________________________________
%                       Observation log  2M0310AB
%-------------------------------------------------------------
\begin{table*}[t!]
\centering
\caption{Observation log of high-angular resolution imaging of  2MASS\,031059+164815\,AB}
\label{ObsLog_2M0310}
\centering
\begin{tabular}{c c c c c c}     % 5 columns
\noalign{\smallskip}
\hline\hline
\noalign{\smallskip}
Date & Telescope/Instrument & Filter & Exp.\,time & Seeing\,$^{\mathit{a}}$ & Strehl ratio\\ 
 & & & [sec]  &  [ $\arcsec$ ] & [ \% ]   \\
\noalign{\smallskip}
\hline
\noalign{\smallskip}
   24\,/\,09\,/\,2004 & HST/NIC1 & F108N & 2560 & \\
   & & F113N & 2816 &  \\
   \noalign{\smallskip}
    05\,/\,11\,/\,2007 & VLT/NACO& H & 14 x 60  & 1.02\,-\,1.19 & 21.9\,-\,43.8\\
    & & K$_\mathrm{S}$   & 14 x 60  & 0.99\,-\,1.07 & 42.2\,-\,68.0 \\
 \noalign{\smallskip}
\hline
\end{tabular}
\begin{list}{}{}
\item[$^{\mathit{a}}$] Site seeing measured by the differential image motion method (DIMM) in \emph{V}-\,band at zenith
\end{list}
\end{table*}
%___________________________________________________
  
Another very peculiar, yet unexplained observational feature is the remarkable brightening in the \emph{Z}/\emph{Y} ($\sim$ 0.9\,-\,1.1 $\mu$m) and \emph{J} ($\sim$ 1.2\,-\,1.3 $\mu$m) bands of up to $\Delta\,M_{J} \sim$\,1\,mag for the early- to mid-type T dwarfs. This so-called \emph{J}\,-\,band ``bump" (\citealt{Dahn, Tinney03, Vrba04}) indicates a significant flux redistribution at  almost constant luminosity. In the following, high-resolution imaging surveys revealed a binary frequency among the L\,/\,T transition objects almost twice as high as in earlier or later type brown dwarfs \citep{Burgasser06_1}. In a first attempt, it was suggested that the ``bump" might be artificially enhanced by systems appearing overluminous due to binarity (``crypto-binarity") and that  the integrated light of an L\,+\,T dwarf system could mimic the spectral characteristics of an early\,-\,type T dwarf (\citealt{Burrows06, Liu06, Burgasser06_1}). However, recent discoveries suggest that at least a fraction of the observed \emph{J}\,-\,band brightening is intrinsic to the atmospheres of early- to mid\,-\,type T dwarfs as they cool. Resolved high-resolution photometry revealed in four L\,/\,T dwarf binary systems a 1.0\,-\,1.3 $\mu$m flux reversal, with the T dwarf secondary being brighter than the late L or early T dwarf primary in this wavelength regime (\object{2MASS J17281150+3948593}\,AB, \citealt{Gizis03}; \object{SDSS J102109.69-030420.1}\,AB, \citealt{Burgasser06_1}; \object{SDSS J153417.05+161546.1}\,AB, \citealt{Liu06}; \object{2MASS J14044941-3159329}\,AB, \citealt{Looper08_2}). In addition, a comparison with absolute \emph{J}\,-\,band magnitudes of other resolved binary components having a spectral type  of T1\,-\,T5 (e.\,g. $\epsilon$\,Indi\,Ba or SDSS\,J042348-041403\,B) shows that they are still  $\sim$ 0.5 mag brighter than the latest L dwarfs (\citealt{Burgasser06_1, Looper08_2}). These findings imply that the brightening across the L\,/\,T transition is a real effect, since it also affects binaries which are assumed to be coeval systems with common age and metallicity.
Therefore, further discoveries and high-resolution observations of L\,/\,T transition binaries will play an important role.  An extended sample of  L\,+\,T dwarf binaries should provide independent crucial information on these issues. The comparison of the observed properties with theoretical models (e.\,g.\,\citealt{Baraffe03, Burrows03, Saumon08}) will then help to reveal the physical mechanism that drives the transition from dusty L dwarfs to dust-free T dwarfs.

One such important newly-resolved binary is \object{2MASS J03105986+1648155} (hereafter 2M0310+1648), whose likely coeval components straddle the L\,/\,T transition. 2M0310+1648 was originally discovered by \citet{Kirk00} in the \emph{Two Micron All Sky Survey} (2MASS) database. It was classified with a spectral type L8 in the optical and the presence of lithium implied a mass $M \le$ 60\,$M_\mathrm{Jup}$, confirming its brown dwarf nature. The first near-IR spectroscopic observations revealed some spectral discrepancy compared to other late-type L dwarfs in terms of a significantly depressed \emph{K} -\,band spectrum starting around 2.2\,$\mu$m, interpreted to be caused by collision-induced H$_2$ absorption \citep{Reid01_2}. By contrast, \citet{Nakaj01} attributed this flux suppression to methane rather than H$_2$, which would also explain  the very weak CO band head at 2.3\,$\mu$m and indicate that much of the carbon is in CH$_4$. Finally, \citet{Geballe02}, with their new classification scheme slightly revised the spectral type of 2M0310+1648 to L9 in the near-IR. 
In September 2004, 2M0310+1648 was resolved as an almost equally bright binary system during our own \emph{Hubble Space Telescope} (HST) NICMOS survey (\citealt{Stumpf05}; Stumpf et al. submitted). As an L\,/\,T transition binary it added up to the already apparently higher binary fraction in the transition regime and provides an important testbed to derive information about the underlying physical and chemical processes in this transition regime. Therefore a monitoring program including resolved photometry and spectroscopy was started and the first results are presented in this paper.
  
%____________________________________________________________
\section{Observations and Data reduction}
\subsection{HST/NICMOS}
\label{2M0310_HST_astrometry}
%_________________________________________________________________
%_______________________________________________________________
%                       2M0310-AB   HST and PARSEC/NACO
%------------------------------------------------------------------------------------
  \begin{figure*}[t!]
  \centering
    \setlength{\unitlength}{1cm}
   \begin{minipage}[t]{4.7cm} 
      \framebox(4.7,4.7){ \includegraphics[width=\textwidth]{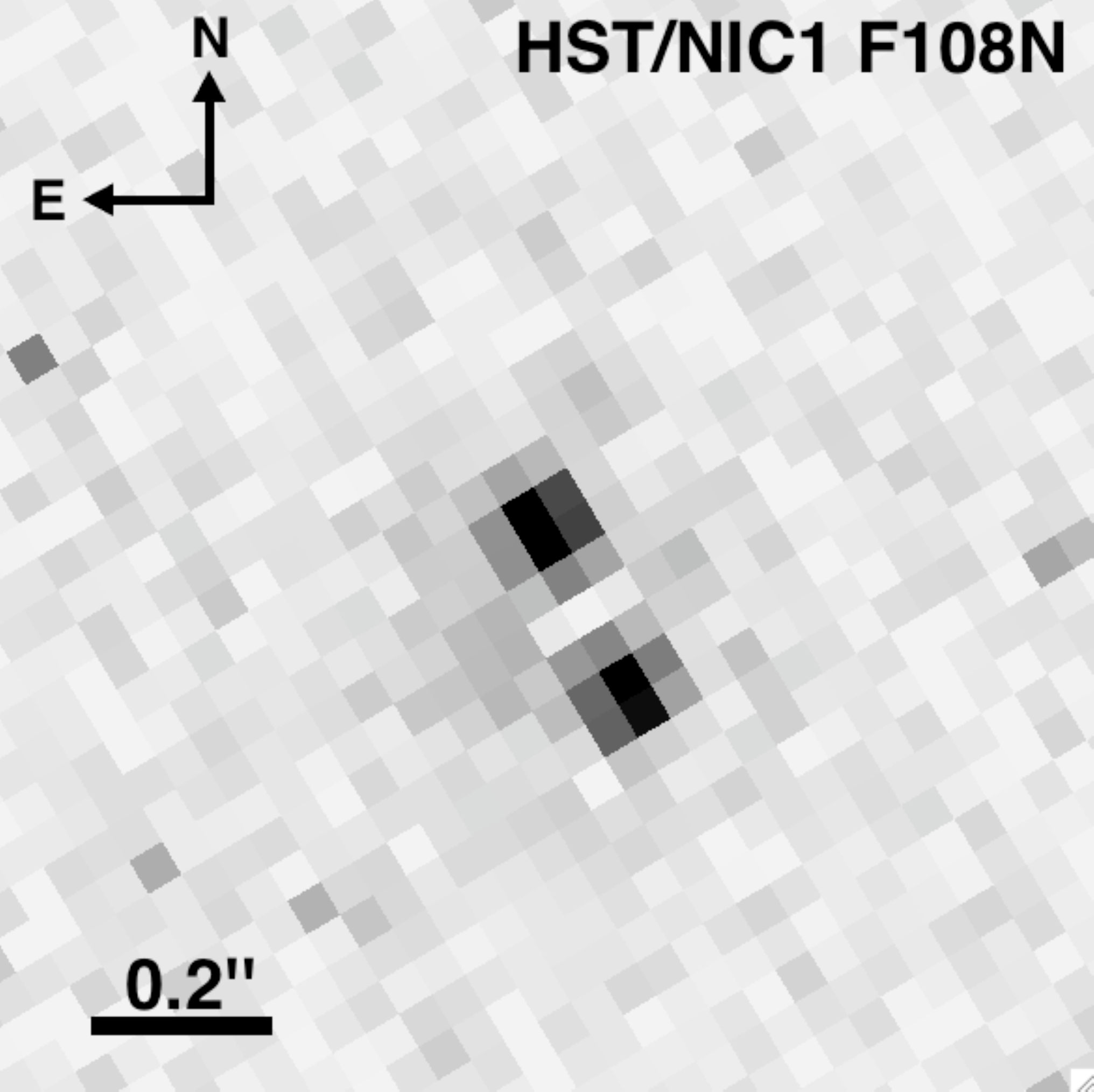}}\vspace{-0.3cm}
        \begin{center}
         24\,/\,09\,/\,2004
       \end{center}
     \end{minipage}\hspace{0.15cm}
    \begin{minipage}[t]{4.7cm} 
         \framebox(4.7,4.7){\includegraphics[width=\textwidth]{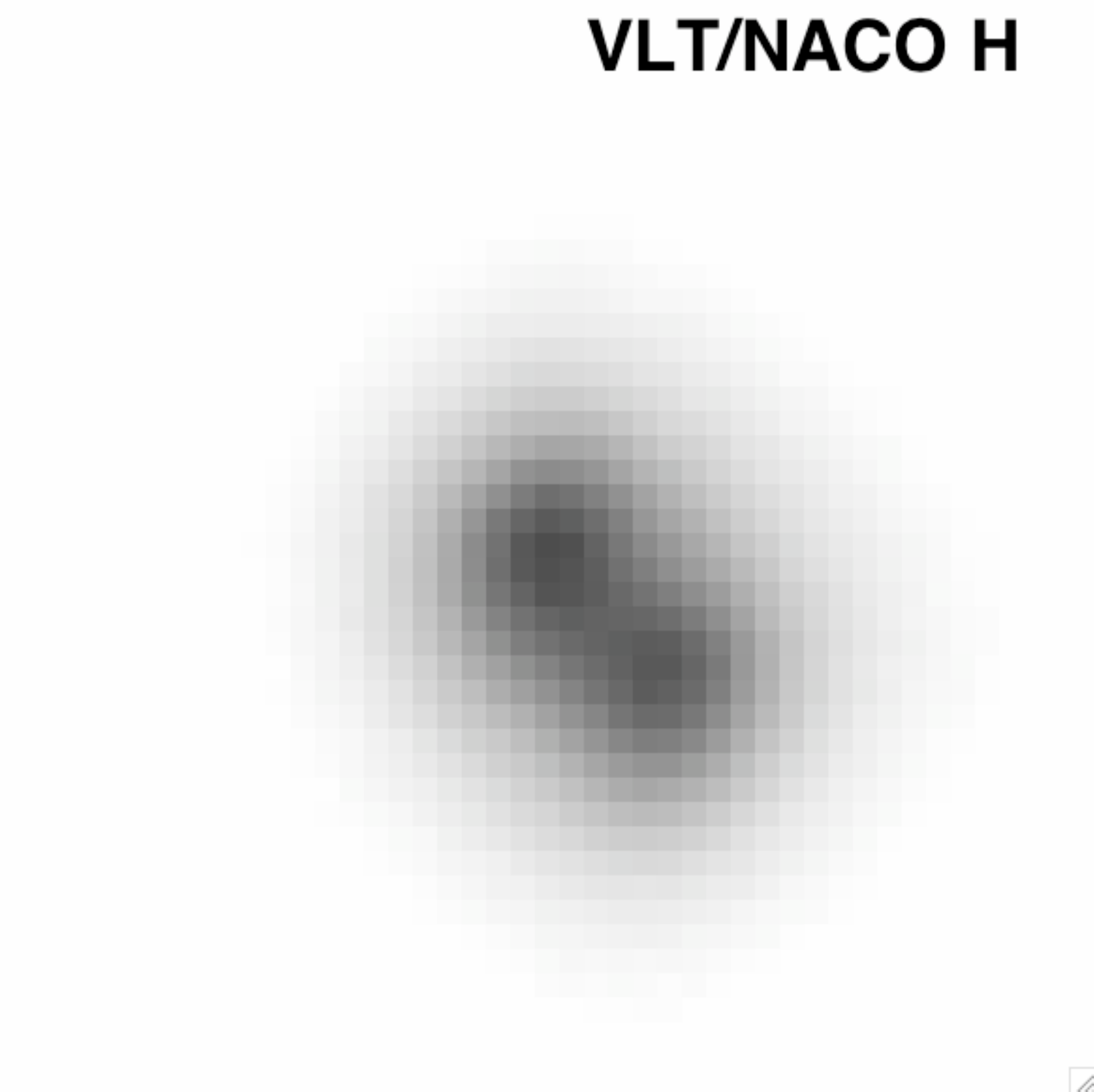}}\vspace{-0.3cm}
        \begin{center}
         05\,/\,11\,/\,2007
       \end{center}
     \end{minipage}\hspace{0.15cm}
    \begin{minipage}[t]{4.7cm}
         \framebox(4.7,4.7){\includegraphics[width=\textwidth]{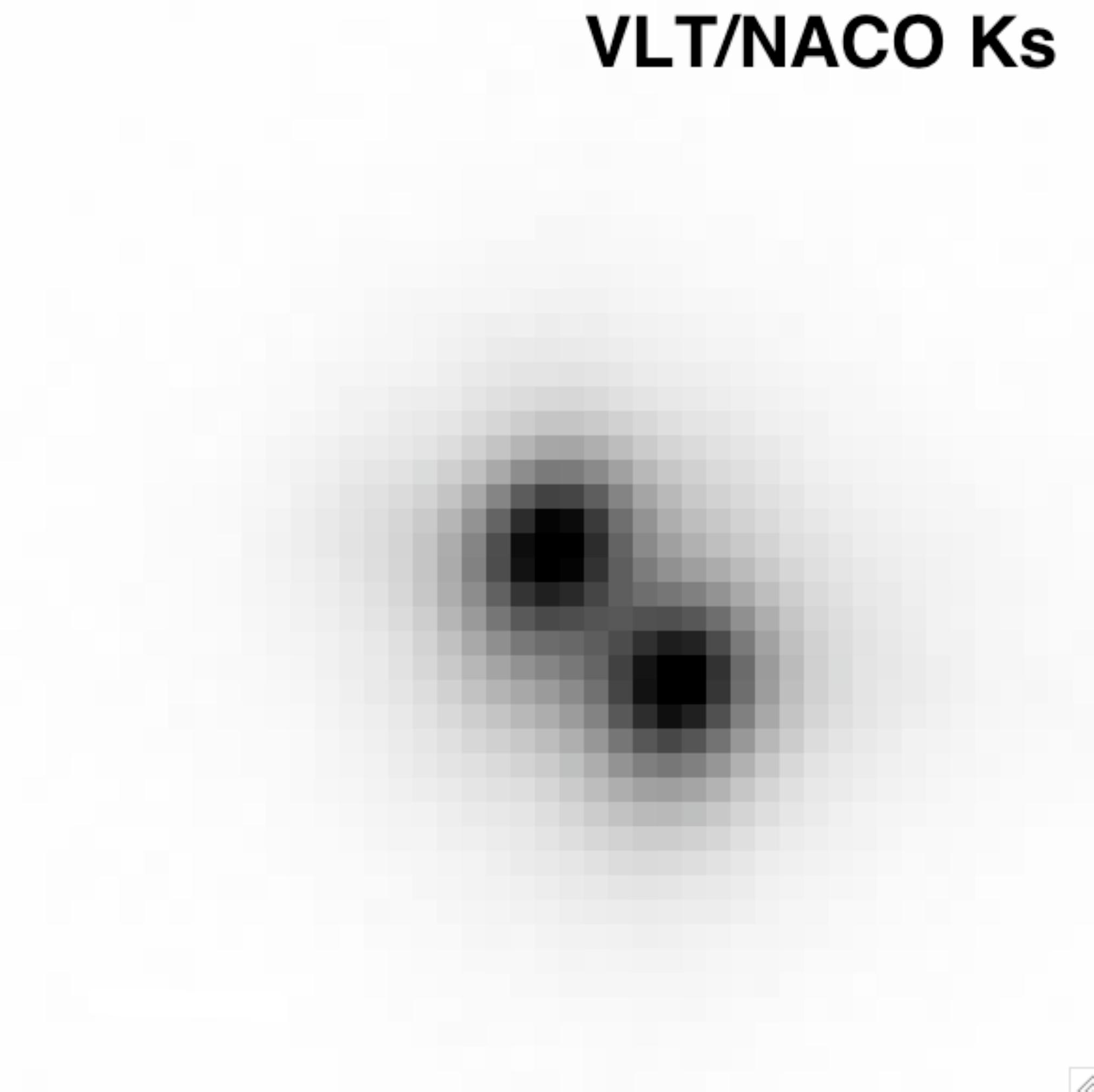}}\vspace{-0.3cm}
        \begin{center}
         05\,/\,11\,/\,2007
       \end{center}
     \end{minipage}\hspace{0.67cm}\vspace{0.3cm}
  \caption{ \footnotesize{Images of 2M0310+1648\,AB obtained with HST/NICMOS and VLT/NACO including PARSEC. The orientation and scale are the same for all images. 2M0310+1648\,A is the north east component while 2M0310+1648\,B is the component to the south west.}}
   \label{2M0310_HST_NACO}
  \end{figure*}
%_____________________________________________________________
%

The HST observations of 2M0310+1648\,AB were obtained as part of the spectral differential imaging program GO 10208 (PI: W. Brandner), targeting 12 isolated L dwarfs with no known companions so far. They were executed on September 24, 2004 with the NICMOS1 (NIC1) camera, providing a field of view (FoV) of 11$\arcsec$\,x\,11$\arcsec$ with a pixel scale of 0.0432$\arcsec$, in the two narrow band filters F108N and F113N. Four different images at two detector positions were acquired in MULTIACCUM mode for each filter, resulting in a total integration time of 2560\,s and 2816\,s, respectively (see Table~\ref{ObsLog_2M0310}).

The HST data analysis of 2M0310+1648\,AB is based on pipeline reduced frames as provided by the HST archive. For the aperture photometry the IRAF \emph{phot} routine in the \emph{apphot} package was used to derive the magnitudes of the resolved components. Due to the small separation between the components an aperture size of 2 pixel was used and corrected to 11.5 pixel using an aperture scaling factor derived from TinyTim\footnote{http://www.stsci.edu/software/tinytim/tinytim.html} PSF simulation. The results were thereafter corrected to a nominal infinite aperture and  transformed into flux using the most recent photometric keyword-value as provided by the STScI webpage\footnote{http://www.stsci.edu/hst/nicmos/performance/photometry/postncs\_keywords.html}. Finally, the flux was converted into the Vega photomertic system using the flux zero points of 1937.0 and 1820.9 Jy for the F108N and F113N filters, respectively. The individual magnitudes from the dithered exposures were then averaged to derive a single photometric measurement for each filter. 

For the astrometric measurements we used the IDL\,-\,based simultaneous PSF\,-\,fitting algorithm from \citet{Bouy03}, adapted to our HST/NIC1 data. For a better error estimation we used a library of 6 different PSFs: four theoretical PSFs considering different focus settings (simulating telescope defocus of up to 10\,$\mu$m due to HST ``breathing") and two natural PSFs obtained during previous observations of similar objects of the same program (namely 2MASSI\,082519+211552  and 2MASSW\,152322+301456, with spectral types of L7.5 and L8, respectively). The precise separation, position angle (PA) and flux ratio were measured separately for each of the 4 images per filter. Finally the results were averaged and the uncertainties were calculated from the standard deviation.

%___________________________________________________________________
\subsection{VLT/NACO with PARSEC}
As part of our MPIA guaranteed time observations (GTO) for the ESO/VLT sodium Laser Guide Star (LGS) system PARSEC \citep{Rabien04, Bonaccini06} we started a monitoring program for a sample of brown dwarf binary systems with spectral types from early L to late T. These observations include photometry and spectroscopy to better constrain the physical properties such as luminosity, colors, spectral types and effective temperature (T$_{\mathrm{eff}}$) of the individual components. 

So far, we have obtained follow-up imaging observations  for 2M0310+1648\,AB with the AO system NACO including PARSEC in the \emph{H} (1.65 $\mu$m) and \emph{K$_\mathrm{S}$} (2.15 $\mu$m) broad-band filters. The observations were carried out in service mode on November 4, 2007 with the CONICA S27 camera, providing a FoV of 28$\arcsec$\,x\,28$\arcsec$ and a pixel scale of 0.0271$\arcsec$. The wavefront sensing was performed on the LGS using the VIS dichroic for the observation and the necessary reference star for Tip/Tilt correction was chosen from the GSC-II (V\,2.2.01). The star N33133125783 has \emph{V}\,=\,17.21 mag and is 46.14$\arcsec$\, away from 2M0310+1648\,AB. Fourteen images were obtained in each filter, executed in a 7 point dither pattern to allow for cosmic ray and bad pixel correction, resulting in 840\,s of total integration time. The observations were performed during clear sky conditions, but wind shake of the telescope and highly variable seeing conditions (between 0.99$\arcsec$ and 1.19$\arcsec$ FWHM), which were far worse than the requested constraint of 0.6$\arcsec$ FWHM, significantly degraded the AO performance especially in \emph{H}\,-band.

The standard image processing included flat-fielding, dark and sky subtraction and bad pixel correction. The final average combination was accomplished using the recommended Eclipse \emph{jitter} \citep{Devil} software package. Since no separate reference PSF star was observed, the simultaneous PSF-fitting algorithm mentioned above from \citet{Bouy03} could not be applied. Therefore, a new IDL fitting algorithm was implemented that fits a system created from two asymetric Moffat PSFs and provides the separation, PA and flux ratio of the two components. The only constraint for this procedure is a similar shape of the observed component PSFs which worked very well for 2M0310+1648\,AB due to the small separation of the binary components. To determine the statistical error of this fit, the algorithm was also applied to each  individually reduced image. All results per filter were averaged and the error calculated from the standard deviation. Figure~\ref{2M0310_HST_NACO} displays the HST data, as well as the final reduced VLT/NACO + PARSEC images in the \emph{H}- and \emph{K$_\mathrm{S}$}\,-\,band.

%__________________________________________________________________

\section{Results}
\subsection{Resolved Photometry and Spectral Types}
%
%                        Photometric properties 2M0310 AB
%-------------------------------------------------------------
\begin{table*}[t!]
\centering
\caption{Resolved component properties of 2MASS\,0310+1648\,AB}
\label{Photometry_2M0310}
\centering
\begin{tabular}{l c c c c}     % 4 columns
\noalign{\smallskip}
\hline\hline
\noalign{\medskip}
Property & flux ratio  &$\Delta$ mag & 2M0310+1648\,A & 2M0310+1648\,B \\
\noalign{\smallskip}
\hline
\noalign{\smallskip}
\noalign{\smallskip}
F108N  &  0.892\,$\pm$\,0.008  & 0.12\,$\pm$\,0.02   & 17.42\,$\pm$\,0.07 & 17.54\,$\pm$\,0.08\\ 
F113N  &  0.891\,$\pm$\,0.008  & 0.12\,$\pm$\,0.02 &17.51\,$\pm$\,0.06 & 17.63\,$\pm$\,0.05 \\[0.5ex]
\emph{J}\,$^{\mathit{*}}$ & 0.908\,$\pm$\,0.020  & 0.10\,$\pm$\,0.06 & 16.73\,$\pm$\,0.11  & 16.83\,$\pm$\,0.13 \\ 
\emph{H}  & 0.947\,$\pm$\,0.012 & 0.059\,$\pm$\,0.013 & 15.66\,$\pm$\,0.08  & 15.71\,$\pm$\,0.08 \\ 
\emph{K$_\mathrm{S}$} & 1.008\,$\pm$\,0.005 &  -0.009\,$\pm$\,0.005  &15.07\,$\pm$\,0.07  & 15.06\,$\pm$\,0.07 \\ [0.5ex]
\emph{J\,--\,H}  &  &  &1.07\,$\pm$\,0.14  &  1.12\,$\pm$\,0.15\\
\emph{H\,--\,K$_\mathrm{S}$}  &  & &0.59\,$\pm$\,0.11 & 0.65\,$\pm$\,0.11 \\
\emph{J\,--\,K$_\mathrm{S}$}  & & &  1.66\,$\pm$\,0.13& 1.77\,$\pm$\,0.15\\[1.0ex]
\noalign{\smallskip}
\hline
\end{tabular}
\begin{list}{}{} 
\item[$^{\mathit{*}}$] the resolved \emph{J}-\,band magnitudes are based on an estimated flux ratio due to the lack of observations (see description in the text).
\end{list}
\end{table*}
%% ________________________________________________

The  magnitudes of the components in the HST F108N and F113N filters are listed in Table~\ref{Photometry_2M0310}. A comparison of the calculated flux ratios from these results with the flux ratios directly derived during the PSF fitting shows a very good agreement within the uncertainties. The slightly fainter magnitudes of both components in the F113N filter compared to those in the F108N filter are real and caused by the increasing water absorption in late L spectral types at this narrow wavelength band. 

The individual component \emph{H} and \emph{K$_\mathrm{S}$} magnitudes were determined from the measured flux ratios in these filters and the published photometry of the unresolved 2M0310+1648 on the 2MASS system \citep{Cutri}. Even though the 2MASS and VLT/NACO near-IR filters are not exactly identical we did not apply any correction factor, given that the spectral energy distributions (SEDs) of the 2M0310+1648\,AB components are so similar that the flux ratio should not be significantly affected by the difference in these two photometric systems.
The final errors in the photometry include the uncertainties of the unresolved 2MASS magnitudes and the determined flux ratios. Due to the lack of \emph{J}\,-\,band observations, the magnitude difference between the two components in this wavelength regime had to be estimated. The SED of L dwarfs is comparatively flat between 1.06\,$\mu$m and 1.15\,$\mu$m as well as throughout the \emph{J}\,-\,band (see \citealp{McLean03}). Thus, the flux ratios between different L spectral types show no significant trend. In addition, both components are almost equally bright suggesting a very similar flux ratio in the F108N and \emph{J} filter. A more quantitative determination was achieved by the convolution of the SED of late L dwarf synthetic spectra with the corresponding filter curves. The result yielded a $\Delta$\,mag correction of 0.02 mag between these two filters. 

Table~\ref{Photometry_2M0310} lists the individual magnitudes and the resulting colors. Within the errors, the two components are equal in magnitude, implying that 2M0310+1648\,AB is likely to be a near equal mass binary (\emph{q} $\sim$\,1). 
However, a closer look at the flux ratios\footnote{\, The flux ratios have smaller and more accurate errors, since the values were derived during direct measurement of the ratio.}  (f$_\mathrm{B}$\,/\,f$_\mathrm{A}$) reveals a steadily decreasing brightness difference between the A and B component from the \emph{Y} (F108N and F113N) to the \emph{K$_\mathrm{S}$}\,-\,band. In the \emph{K$_\mathrm{S}$}\,-\,band a flux reversal even occurs between the components. This is puzzling, since one would not expect the inversion of the brightness ratio if both dwarfs were equal. A similar flux reversal, although in the \emph{J}\,-\,band, has been detected in the four other L\,/\,T transition binaries described in $\S$\,\ref{LT_transition_intro}. Three of these previously detected reversal binaries (SDSS\,1021-0304\,AB, SDSS\,1534+1615\,AB, 2M1404-3159\,AB) are composed of a T1\,-\,T1.5 primary and a T5\,-\,T5.5 secondary, where the secondary is brighter in \emph{J} ($\Delta$\,\emph{J} $\sim$ 0.04\,-\,0.54 mag in the MKO system) but significantly fainter in \emph{H} and \emph{K$_\mathrm{S}$}. First, 2M0310+1648\,AB seems to be different, since the flux reversal appears in the \emph{K$_\mathrm{S}$}\,-\,band. However, if one would consider that the assumed primary 2M0310+1648\,A\footnote{\, The history of naming the actual A component as the primary, arose from the first resolved photometry derived with HST, where "A" was the brighter object.} is actually of slightly later spectral type than the assumed secondary 2M0310+1648\,B, the later-type (A) component would be also notably brighter in \emph{Y} and \emph{J}\,-\,band and fainter in \emph{K$_\mathrm{S}$}. A possible explanation for the still slightly brighter flux in \emph{H}\,-\,band could be that both components have much closer spectral types, thus very similar SEDs with similar strong H$_2$O and CH$_4$ absorption, in contrast to the T1\,+\,T5 binaries mentioned above. Therefore, the inversion might take place in a more continuous way.

\begin{figure*}[t!]
  \centering
       \includegraphics[width=10.6cm]{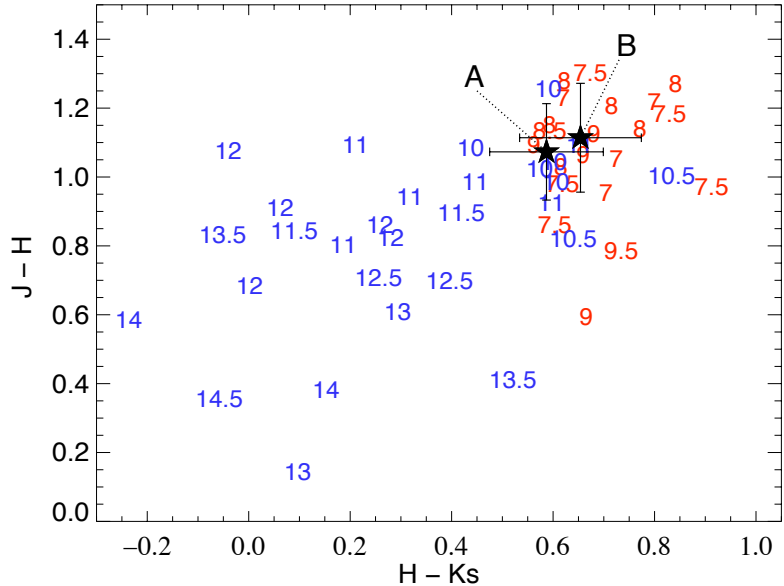}
  \caption{\footnotesize{Near-IR color - color diagram of 61 known L7\,-\,T4.5 dwarfs from the \emph{DwarfArchive} with a near-IR spectral type uncertainty $\le$1 subtype and known binary systems excluded. The L dwarfs are shown in red while the T dwarfs are shown in blue. The numbers indicate the individual spectral types from 7\,=\,L7 to 14.5\,=\,T4.5. The resolved colors of 2M0310+1648\,A and B are plotted as stars with corresponding error bars.}}
   \label{color_color_2M0310}
  \end{figure*}

To derive more information on the individual spectral types, they are assumed to be identical in a first approach, since both components are almost equally bright and thus should not have significantly different spectra compared to the unresolved 2M0310+1648 spectrum which has an assigned spectral type of L9. Additionally, while the flux of both components drops in the F113N filter due to water absorption, their flux ratio remains the same as in F108N, indicating the same strength of absorption in each component and thus the same spectral type.
To check these estimates, the resolved \emph{JHKs} colors of 2M0310+1648\,AB are compared to those of 61 known L7\,-\,T4.5 dwarfs from the \emph{Dwarf Archive}\footnote{\, http://www.DwarfArchive.org} (provided in the 2MASS photometric system), excluding any known binaries. The color\,-\,color diagram in Figure~\ref{color_color_2M0310} shows that both components have colors coincident with a cluster of late L dwarfs (L7\,-\,L9) and very early T dwarfs (T0\,-\,T1), supporting the assumption that 2M0310+1648\,A and B have a spectral type $\sim$ L9. At the same time, the color composition illustrates the peculiar redder colors of component B compared to component A, indicating the possibility that 2M0310+1648\,A is of slightly later spectral type, possibly T0. In fact, a comparison of 2M0310+1648\,A with the T0 standard \object{SDSS J120747.17+024424.8} \citep{Tinney05, Burgasser06_3} reveals a very good agreement of the  colors within the uncertainties (SDSS\,1207: \emph{J\,--\,H}\,=\,1.02\,$\pm$\,0.09, \emph{H\,--\,K$_\mathrm{S}$}\,=\,0.57\,$\pm$\,0.09,  \emph{J\,--\,K$_\mathrm{S}$}\,=\,1.59\,$\pm$\,0.09). However, a wider spread of colors for the same spectral type is not uncommon for dwarfs in the L\,/\,T transition (see, e.\,g.\,\citealp{Knapp}). 

Hence, only resolved spectroscopy can provide unambiguous spectral type determinations, which will further help to explain the physical mechanisms in the L/T transition. For the present, the assumption that both components have the same spectral type of L9\,$\pm$\,1 in the near-IR will be retained until spectroscopic results are obtained.
%
%                        Color - Color-Diagramm 2M0310
%-------------------------------------------------------------
%___________________________________________

%
\subsection{New photometric distance}
Since 2M0310+1648 has no trigonometric parallax determination so far, an assigned photometric distance of 20\,pc (\citealt{Kirk00}) has commonly been used for the unresolved system. To correct this value for the bias introduced by the multiplicity of the system, the individual component magnitudes were compared to the absolute magnitude vs.\,Spectral Type (SpT) relation from \citet{Looper08_2}\footnote{\, In contrast to the \citet{Burgasser07_2} relation, this relation is based on 2MASS photometry, thus not introducing additional errors due to the conversion from one photometric system into another.}. With an assumed near-IR spectral type of L9 for each component this $M_\mathrm{JHK_S}$ -- SpT relation gives $M_{H}$\,=\,13.67\,$\pm$\,0.29 mag and $M_{K_S}$\,=\,13.04\,$\pm$\,0.33 mag. This yields distances of $\sim$\,24.9 and $\sim$\,25.4 pc, respectively, for component A and $\sim$\,25.6 and $\sim$\,25.3 pc, respectively, for component B. With these results a mean distance of 25\,$\pm$\,4 pc is assigned, including the uncertainties in the photometric magnitudes and the rms error in the spectral type relation.

\subsection{Orbit estimates}
Table~\ref{OrbPar_2M0310} lists the measured separations and position angles for the 2M0310+1648\,AB system obtained from two epochs separated by $\sim$ 3 years. During that time, the position angle changed by 15.5\degr\,$\pm$ 0.8\degr\,while the separation increased only slightly from 204.3\,$\pm$\,0.4\,mas to 210.8 $\pm$ 1.8\,mas. For a first estimation of the orbital parameters, a face-on circular orbit was assumed, since the separation between the components did not change significantly. Accordingly, an average value of 207.6\,$\pm$\,1.8\,mas was used in the following calculations. This resulted in the approximation that the semi-major axis corresponds to a projected separation of 5.2\,$\pm$\,0.8 AU (at a distance of 25 pc) with the uncertainty being dominated by the distance estimate. With an orbital period of 72\,$\pm$\,4 years calculated from the fractional change in PA, and using Kepler's third law, this finally yields a first estimate for the total system mass of $\approx$ 30\,$M_\mathrm{Jup}$. This implies a relatively low-mass binary brown dwarf system with both components having masses close to the brown dwarf/planetary mass boundary. 
%
%                        Orbit parameters 2M0310
%-------------------------------------------------------------
\begin{table}[t!]
\centering
\caption{Orbital parameters for the 2MASS0310+1648\,AB system}
\label{OrbPar_2M0310}
\centering
\begin{tabular}{l l c c }     % 3 columns
\noalign{\smallskip}
\hline\hline
\noalign{\medskip}
& & HST/NIC1 &VLT/NACO\\
\raisebox{1.5ex}[1.5ex]{Parameter} & & (24\,/\,09\,/\,2004) & (05\,/\,11\,/\,2007)\\
\noalign{\medskip}
\hline
\noalign{\smallskip}
\noalign{\smallskip}
Separation & $\rho$ [mas] &  204.3 $\pm$ 0.4 & 210.8 $\pm$ 1.8 \\[0.5ex]
Position Angle & $\theta$ [deg]   &  206.4 $\pm$ 0.1 &  221.9 $\pm$ 0.8 \\[1ex]
estimated distance & $d$ [pc] &  \multicolumn{2}{c}{25 $\pm$ 4}\\ [0.5ex]
Semi-major axis & $a$ [AU]   & \multicolumn{2}{c}{\hspace{-2.5ex}$\ge$\, 5.2 $\pm$ 0.8}\\ [0.5ex]
Orbital period & $P$ [years]   & \multicolumn{2}{c}{72 $\pm$ 4}\\[0.5ex]
System mass & $M_{tot}$ [$M_\mathrm{Jup}$]  & \multicolumn{2}{c}{\hspace{-1.5ex}$\ge$\,  30}\\ 
\noalign{\smallskip}
\hline
\end{tabular}
\end{table}
%% ________________________________________________

However, as the projected separation was assumed to be equal with the true semi-major axis, the mass prediction can only be a lower limit. Depending on the orientation in space (different perspectives on inclination \emph{i} and eccentricity \emph{e}), the true separation might be larger and\,/\,or the observations may have been obtained very close to the periastron passage, resulting in an increase of the total system mass. \citet{Fischer1992} showed that on average the true semi-major axis for binaries is about 1.26 times larger than the observed separation. Correcting with this statistical factor, the true semi-major axis for 2M0310+1648\,AB can be estimated as $a$ = 1.26\,$\ast$\,$\rho$\, $\approx$ 6.6\,AU. Using the same orbital\- period as before implies an almost doubled total system mass approximation of $\approx$ 60\,$M_\mathrm{Jup}$.
This discrepancy shows the need for further astrometric observations to better determine the orbital parameters and to clarify if the orbit is really seen face on or at a different inclination.

%__________________________________________________________________
\section{Conclusions}
HST/NICMOS imaging in the F108N and F113N filters revealed the binary nature of another very interesting L/T transition brown dwarf: 2M0310+1648\,AB. In the following, second epoch astrometry and first resolved high-resolution photometry in the \emph{H}\,- and \emph{K$_\mathrm{S}$}\,-\,bands were obtained with VLT/NACO and its new LGS AO system PARSEC.

The two epochs of astrometric measurements spanning $\sim$ 3 years allowed for first rough orbital parameter estimations. Due to a non-significant change in the separation,  a face-on circular orbit was assumed, yielding  an orbital period of 72\,$\pm$\,4 years. Depending on the assumed semi-major axis, Kepler's third law yielded a first total system mass estimate of $\sim$ 30\,-\,60\,$M_\mathrm{Jup}$, placing the individual component masses at the lower end of the brown dwarf regime. The first orbital period estimate of $\sim$ 72 years does not suggest the possibility for a meaningful dynamical mass determination on a short time scale. Nevertheless, follow-up observations in the next years will allow us to derive more accurate information on the orbital elements and hence the true orientation of the system in space, as well as the true orbital period. This will finally enable us to better constrain the total system mass. 

The derived photometry revealed a very intriguing property of 2M0310+1648\,AB. The component fluxes show an unexpected decrease in brightness difference with increasing wavelength, resulting in a marginal flux reversal in the \emph{K$_\mathrm{S}$}\,-\,band. An additional comparison of the component colors obtained reveals a redder color of the B component. These results indicate that the designated primary component 2M0310+1648\,A might actually be of slightly later spectral type than 2M0310+1648\,B. This could at least partly explain the observed flux reversal as part of the \emph{J}\,-\,band brightening of early- to mid-type T dwarfs, but a full explanation for the true nature of the reversal is still owing. Upcoming spatially resolved spectroscopic observations with VLT/SINFONI and the PARSEC AO system will allow a precise spectral type determination and an investigation of the underlying spectral morphologies. If it turns out that 2M0310+1648\,A is really of later spectral type than 2M0310+1648\,B, the system  would add up to the currently small sample of flux reversal binaries. Additionally, 2M0310+1648\,AB would be the first binary with a secondary showing the \emph{J}\,-\,band brightening already at the very late\,-\,L (L9) or very early-T (T0) dwarf stage rather than at a T1.5 spectral type or later. This would  challenge the existing theoretical models even more.

In future work, the likely coeval system 2M0310+1648\,AB will serve as a very important benchmark object in the L\,/\,T transition. Further high-resolution observations will provide  an improved understanding of and new insights into the challenging picture of this still poorly understood, yet remarkable evolutionary phase of brown dwarfs.

%__________________________________________________________________
\begin{acknowledgements}
M.B.Stumpf and W. Brandner acknowledge support by the \emph{DLR Verbundforschung} project numbers 50 OR 0401 and 50 OR 0902. We are grateful to Tricia Royle at STScI, Lowell Tacconi-Garman at ESO and the staff of ESO/Paranal for their great and efficient support before and during observations.
We would like to thank the anonymous referee for the constructive comments, which helped to improve the paper.
This research has benefitted from the M, L and T dwarf compendium housed at DwarfArchives.org and maintained by Chris Gelino, Davy Kirkpatrick and Adam Burgasser. This research has made use of the SIMBAD database, operated at CDS, Strasbourg, France
\end{acknowledgements}

\bibliographystyle{aa}
\bibliography{literature.bib}

\small
\begin{thebibliography}{34}
\expandafter\ifx\csname natexlab\endcsname\relax\def\natexlab#1{#1}\fi

\bibitem[{{Ackerman} \& {Marley}(2001)}]{Ackerman01}
{Ackerman}, A.~S. \& {Marley}, M.~S. 2001, \apj, 556, 872

\bibitem[{{Baraffe} {et~al.}(2003){Baraffe}, {Chabrier}, {Barman}, {Allard}, \&
  {Hauschildt}}]{Baraffe03}
{Baraffe}, I., {Chabrier}, G., {Barman}, T.~S., {Allard}, F., \& {Hauschildt},
  P.~H. 2003, \aap, 402, 701

\bibitem[{{Bonaccini Calia} {et~al.}(2006){Bonaccini Calia}, {Allaert},
  {Alvarez}, {Araujo Hauck}, {Avila}, {Bendek}, {Buzzoni}, {Comin}, {Cullum},
  {Davies}, {Dimmler}, {Guidolin}, {Hackenberg}, {Hippler}, {Kellner}, {van
  Kesteren}, {Koch}, {Neumann}, {Ott}, {Popovic}, {Pedichini}, {Quattri},
  {Quentin}, {Rabien}, {Silber}, \& {Tapia}}]{Bonaccini06}
{Bonaccini Calia}, D., {Allaert}, E., {Alvarez}, J.~L., {et~al.} 2006, in
  Society of Photo-Optical Instrumentation Engineers (SPIE) Conference Series,
  Vol. 6272, SPIE, 627207

\bibitem[{{Bouy} {et~al.}(2003){Bouy}, {Brandner}, {Mart{\'{\i}}n}, {Delfosse},
  {Allard}, \& {Basri}}]{Bouy03}
{Bouy}, H., {Brandner}, W., {Mart{\'{\i}}n}, E.~L., {et~al.} 2003, \aj, 126,
  1526

\bibitem[{{Burgasser}(2007)}]{Burgasser07_2}
{Burgasser}, A.~J. 2007, \apj, 659, 655

\bibitem[{{Burgasser} {et~al.}(2006{\natexlab{a}}){Burgasser}, {Geballe},
  {Leggett}, {Kirkpatrick}, \& {Golimowski}}]{Burgasser06_3}
{Burgasser}, A.~J., {Geballe}, T.~R., {Leggett}, S.~K., {Kirkpatrick}, J.~D.,
  \& {Golimowski}, D.~A. 2006{\natexlab{a}}, \apj, 637, 1067

\bibitem[{{Burgasser} {et~al.}(2006{\natexlab{b}}){Burgasser}, {Kirkpatrick},
  {Cruz}, {Reid}, {Leggett}, {Liebert}, {Burrows}, \& {Brown}}]{Burgasser06_1}
{Burgasser}, A.~J., {Kirkpatrick}, J.~D., {Cruz}, K.~L., {et~al.}
  2006{\natexlab{b}}, \apjs, 166, 585

\bibitem[{{Burgasser} {et~al.}(2002){Burgasser}, {Marley}, {Ackerman},
  {Saumon}, {Lodders}, {Dahn}, {Harris}, \& {Kirkpatrick}}]{Burgasser02}
{Burgasser}, A.~J., {Marley}, M.~S., {Ackerman}, A.~S., {et~al.} 2002, \apjl,
  571, L151

\bibitem[{{Burrows} {et~al.}(2003){Burrows}, {Sudarsky}, \&
  {Hubbard}}]{Burrows03}
{Burrows}, A., {Sudarsky}, D., \& {Hubbard}, W.~B. 2003, \apj, 594, 545

\bibitem[{{Burrows} {et~al.}(2006){Burrows}, {Sudarsky}, \&
  {Hubeny}}]{Burrows06}
{Burrows}, A., {Sudarsky}, D., \& {Hubeny}, I. 2006, \apj, 640, 1063

\bibitem[{{Cutri} {et~al.}(2003){Cutri}, {Skrutskie}, {van Dyk}, {Beichman},
  {Carpenter}, {Chester}, {Cambresy}, {Evans}, {Fowler}, {Gizis}, {Howard},
  {Huchra}, {Jarrett}, {Kopan}, {Kirkpatrick}, {Light}, {Marsh}, {McCallon},
  {Schneider}, {Stiening}, {Sykes}, {Weinberg}, {Wheaton}, {Wheelock}, \&
  {Zacarias}}]{Cutri}
{Cutri}, R.~M., {Skrutskie}, M.~F., {van Dyk}, S., {et~al.} 2003, {2MASS All
  Sky Catalog of point sources.} (The IRSA 2MASS All-Sky Point Source Catalog,
  NASA/IPAC Infrared Science
  Archive.~http://irsa.ipac.caltech.edu/applications/Gator/)

\bibitem[{{Dahn} {et~al.}(2002){Dahn}, {Harris}, {Vrba}, {Guetter}, {Canzian},
  {Henden}, {Levine}, {Luginbuhl}, {Monet}, {Monet}, {Pier}, {Stone}, {Walker},
  {Burgasser}, {Gizis}, {Kirkpatrick}, {Liebert}, \& {Reid}}]{Dahn}
{Dahn}, C.~C., {Harris}, H.~C., {Vrba}, F.~J., {et~al.} 2002, \aj, 124, 1170

\bibitem[{{Devillard}(1997)}]{Devil}
{Devillard}, N. 1997, The ESO Messenger, 87, 19

\bibitem[{{Fischer} \& {Marcy}(1992)}]{Fischer1992}
{Fischer}, D.~A. \& {Marcy}, G.~W. 1992, \apj, 396, 178

\bibitem[{{Folkes} {et~al.}(2007){Folkes}, {Pinfield}, {Kendall}, \&
  {Jones}}]{Folkes07}
{Folkes}, S.~L., {Pinfield}, D.~J., {Kendall}, T.~R., \& {Jones}, H.~R.~A.
  2007, \mnras, 378, 901

\bibitem[{{Geballe} {et~al.}(2002){Geballe}, {Knapp}, {Leggett}, {Fan},
  {Golimowski}, {Anderson}, {Brinkmann}, {Csabai}, {Gunn}, {Hawley},
  {Hennessy}, {Henry}, {Hill}, {Hindsley}, {Ivezi{\'c}}, {Lupton}, {McDaniel},
  {Munn}, {Narayanan}, {Peng}, {Pier}, {Rockosi}, {Schneider}, {Smith},
  {Strauss}, {Tsvetanov}, {Uomoto}, {York}, \& {Zheng}}]{Geballe02}
{Geballe}, T.~R., {Knapp}, G.~R., {Leggett}, S.~K., {et~al.} 2002, \apj, 564,
  466

\bibitem[{{Gizis} {et~al.}(2003){Gizis}, {Reid}, {Knapp}, {Liebert},
  {Kirkpatrick}, {Koerner}, \& {Burgasser}}]{Gizis03}
{Gizis}, J.~E., {Reid}, I.~N., {Knapp}, G.~R., {et~al.} 2003, \aj, 125, 3302

\bibitem[{{Golimowski} {et~al.}(2004){Golimowski}, {Henry}, {Krist},
  {Dieterich}, {Ford}, {Illingworth}, {Ardila}, {Clampin}, {Franz},
  {Wasserman}, {Benedict}, {McArthur}, \& {Nelan}}]{Goli04_2}
{Golimowski}, D.~A., {Henry}, T.~J., {Krist}, J.~E., {et~al.} 2004, \aj, 128,
  1733

\bibitem[{{Kirkpatrick} {et~al.}(2000){Kirkpatrick}, {Reid}, {Liebert},
  {Gizis}, {Burgasser}, {Monet}, {Dahn}, {Nelson}, \& {Williams}}]{Kirk00}
{Kirkpatrick}, J.~D., {Reid}, I.~N., {Liebert}, J., {et~al.} 2000, \aj, 120,
  447

\bibitem[{{Knapp} {et~al.}(2004){Knapp}, {Leggett}, {Fan}, {Marley}, {Geballe},
  {Golimowski}, {Finkbeiner}, {Gunn}, {Hennawi}, {Ivezi{\'c}}, {Lupton},
  {Schlegel}, {Strauss}, {Tsvetanov}, {Chiu}, {Hoversten}, {Glazebrook},
  {Zheng}, {Hendrickson}, {Williams}, {Uomoto}, {Vrba}, {Henden}, {Luginbuhl},
  {Guetter}, {Munn}, {Canzian}, {Schneider}, \& {Brinkmann}}]{Knapp}
{Knapp}, G.~R., {Leggett}, S.~K., {Fan}, X., {et~al.} 2004, \aj, 127, 3553

\bibitem[{{Liu} {et~al.}(2006){Liu}, {Leggett}, {Golimowski}, {Chiu}, {Fan},
  {Geballe}, {Schneider}, \& {Brinkmann}}]{Liu06}
{Liu}, M.~C., {Leggett}, S.~K., {Golimowski}, D.~A., {et~al.} 2006, \apj, 647,
  1393

\bibitem[{{Looper} {et~al.}(2008){Looper}, {Gelino}, {Burgasser}, \&
  {Kirkpatrick}}]{Looper08_2}
{Looper}, D.~L., {Gelino}, C.~R., {Burgasser}, A.~J., \& {Kirkpatrick}, J.~D.
  2008, \apj, 685, 1183

\bibitem[{{Marley} {et~al.}(2002){Marley}, {Seager}, {Saumon}, {Lodders},
  {Ackerman}, {Freedman}, \& {Fan}}]{Marley02}
{Marley}, M.~S., {Seager}, S., {Saumon}, D., {et~al.} 2002, \apj, 568, 335

\bibitem[{{McLean} {et~al.}(2003){McLean}, {McGovern}, {Burgasser},
  {Kirkpatrick}, {Prato}, \& {Kim}}]{McLean03}
{McLean}, I.~S., {McGovern}, M.~R., {Burgasser}, A.~J., {et~al.} 2003, \apj,
  596, 561

\bibitem[{{Nakajima} {et~al.}(2001){Nakajima}, {Tsuji}, \&
  {Yanagisawa}}]{Nakaj01}
{Nakajima}, T., {Tsuji}, T., \& {Yanagisawa}, K. 2001, \apjl, 561, L119

\bibitem[{{Rabien} {et~al.}(2004){Rabien}, {Davies}, {Ott}, {Li}, {Abuter},
  {Kellner}, \& {Neumann}}]{Rabien04}
{Rabien}, S., {Davies}, R.~I., {Ott}, T., {et~al.} 2004, in Society of
  Photo-Optical Instrumentation Engineers (SPIE) Conference Series, Vol. 5490,
  SPIE Conference Series, ed. D.~{Bonaccini Calia}, B.~L. {Ellerbroek}, \&
  R.~{Ragazzoni}, 981--988

\bibitem[{{Reid} {et~al.}(2001){Reid}, {Burgasser}, {Cruz}, {Kirkpatrick}, \&
  {Gizis}}]{Reid01_2}
{Reid}, I.~N., {Burgasser}, A.~J., {Cruz}, K.~L., {Kirkpatrick}, J.~D., \&
  {Gizis}, J.~E. 2001, \aj, 121, 1710

\bibitem[{{Saumon} \& {Marley}(2008)}]{Saumon08}
{Saumon}, D. \& {Marley}, M.~S. 2008, ArXiv e-prints

\bibitem[{{Stumpf} {et~al.}(2005){Stumpf}, {Brandner}, \& {Henning}}]{Stumpf05}
{Stumpf}, M.~B., {Brandner}, W., \& {Henning}, T. 2005, Protostars and Planets
  V, 8571

\bibitem[{{Tinney} {et~al.}(2003){Tinney}, {Burgasser}, \&
  {Kirkpatrick}}]{Tinney03}
{Tinney}, C.~G., {Burgasser}, A.~J., \& {Kirkpatrick}, J.~D. 2003, \aj, 126,
  975

\bibitem[{{Tinney} {et~al.}(2005){Tinney}, {Burgasser}, {Kirkpatrick}, \&
  {McElwain}}]{Tinney05}
{Tinney}, C.~G., {Burgasser}, A.~J., {Kirkpatrick}, J.~D., \& {McElwain}, M.~W.
  2005, \aj, 130, 2326

\bibitem[{{Tsuji}(2005)}]{Tsuji05}
{Tsuji}, T. 2005, \apj, 621, 1033

\bibitem[{{Tsuji} {et~al.}(1999){Tsuji}, {Ohnaka}, \& {Aoki}}]{Tsuji99}
{Tsuji}, T., {Ohnaka}, K., \& {Aoki}, W. 1999, \apjl, 520, L119

\bibitem[{{Vrba} {et~al.}(2004){Vrba}, {Henden}, {Luginbuhl}, {Guetter},
  {Munn}, {Canzian}, {Burgasser}, {Kirkpatrick}, {Fan}, {Geballe},
  {Golimowski}, {Knapp}, {Leggett}, {Schneider}, \& {Brinkmann}}]{Vrba04}
{Vrba}, F.~J., {Henden}, A.~A., {Luginbuhl}, C.~B., {et~al.} 2004, \aj, 127,
  2948

\end{thebibliography}

\end{document}